\documentclass[9pt,twocolumn,twoside]{osajnl}

\journal{ol} 

\setboolean{shortarticle}{true} 

\ifthenelse{\boolean{shortarticle}}{\colorlet{color2}{color2b}}{\colorlet{color2}{color2}} 

\title{Deuterated silicon nitride photonic devices for broadband optical frequency comb generation}

\author[1,*]{Jeff Chiles}
\author[1]{Nima Nader}
\author[2]{Daniel D. Hickstein}
\author[2]{Su Peng Yu}
\author[2]{Travis Crain Briles}
\author[2]{David Carlson}
\author[2]{Hojoong Jung}
\author[1]{Jeffrey M. Shainline}
\author[2]{Scott Diddams}
\author[2]{Scott B. Papp}
\author[1]{Sae Woo Nam}
\author[1]{Richard P. Mirin}

\affil[1]{Applied Physics Division, National Institute of Standards and Technology, 325 Broadway, Boulder, CO 80305}
\affil[2]{Time and Frequency Division, National Institute of Standards and Technology, 325 Broadway, Boulder, CO 80305}
\affil[*]{Corresponding author: jeffrey.chiles@nist.gov}

\dates{Compiled \today}

\ociscodes{}

\doi{}

\begin{abstract}
We report and characterize low-temperature, plasma-deposited deuterated silicon nitride thin films for nonlinear integrated photonics.  With a peak processing temperature less than 300$^\circ$C, it is back-end compatible with pre-processed CMOS substrates.  We achieve microresonators with a quality factor of up to $1.6\times 10^6 $ at 1552 nm, and $>1.2\times 10^6$ throughout $\lambda = 1510-1600$ nm, without annealing or stress management.  We then demonstrate the immediate utility of this platform in nonlinear photonics by generating a 1 THz free spectral range, 900-nm-bandwidth modulation-instability microresonator Kerr comb and octave-spanning, supercontinuum-broadened spectra.
\end{abstract}

\setboolean{displaycopyright}{false}

\begin{document}

\thispagestyle{fancy}
\maketitle

Integrated photonics provides a versatile platform for the study and exploitation of nonlinear optical phenomena. Technological developments in this area have long been motivated by the quest to realize compact, lightweight and efficient optical frequency converters and synthesizers, as well as broadband and stabilized frequency combs \cite{Kippenberg2011}.  But to achieve such technologies, a suitable material platform is required.  Of the many materials that have been considered, trade-offs exist between the strength of the Kerr nonlinearity, refractive index, dispersion, and linear/nonlinear absorption \cite{Moss2013}.  To date, stoichiometric, low-pressure-chemical-vapor-deposited (LPCVD) silicon nitride (SiN) has been a highly successful material, despite its modest nonlinearity compared to many semiconductor alternatives. The relative ease of realizing low-loss waveguides and its superior transparency window at shorter wavelengths (enabling simple and inexpensive telecom-band pumping) have bolstered its position as the material of choice.  These properties have resulted in numerous developments such as soliton microcombs \cite{Brasch2016}, normal-dispersion microcombs \cite{Liu2014}, ultra-low power optical parametric oscillators \cite{Ji2016}, supercontinuum generation for self-referenced frequency combs at low pulse energies \cite{Carlson2017}, and mid-infrared microcombs \cite{Luke2015}.

These outcomes are possible because of the substantial research into improving the processing and properties of SiN films produced via LPCVD \cite{Luke2013,Pfeiffer:16,Ji2016}.  The major lessons learned include how to avoid crack propagation from the intrinsically high stress of thick films, as well as the minimization of hydrogen content (which normally causes substantial loss due to the N-H overtone absorption peak from 1500 -- 1550 nm \cite{Douglas2016}). All modern LPCVD SiN processes either involve high-temperature anneal and deposition cycles (adding to expense and yield degradation), or suffer from occasional cracks that propagate into device regions.  Furthermore, such films cannot be directly deposited on top of pre-processed CMOS wafers or chips, due to the high temperatures (> 700$^\circ$C) required.  The same limitation applies for various temperature-sensitive photonic platforms such as thin-film LiNbO\textsubscript{3}\cite{Rabiei2013a} on silicon, necessitating complex integration alternatives such as direct bonding \cite{Chang2017}.  These are causes for concern when evaluating the commercial viability of SiN photonics.

Plasma-enhanced chemical vapor deposition (PECVD), on the other hand, is a straightforward and commercially favorable processing technique with many advantages including simple control over film stress, high repeatability, and low temperature requirements ($<$400$^\circ$C).  This has motivated a parallel branch of research into optimizing SiN films deposited at low temperatures \cite{Mao2008,Douglas2016,Shao2016,Ooi2017}.  The predominant limitation for most PECVD SiN films is the loss due to N-H bonds, since the hydrogen content is high in low-temperature, unannealed thin films.  Thus, most studies have focused on reducing the content of N-H bonds in particular.  Still, the worst-case propagation loss throughout the 1500 -- 1600 nm wavelength range has not been reduced below 2.5 dB per cm \cite{Douglas2016}. This is too high to be useful for most photonics applications, since it limits bandwidth, prevents long propagation lengths and increases susceptibility to optical damage.

An effective solution is to employ isotopically substituted precursors during the deposition process to modify the bond energy of the N-H overtone (such as deuterated silane, SiD\textsubscript{4} instead of conventional silane, SiH\textsubscript{4}) \cite{Beyeler2001}.  This has been shown to shift the absorption band from 1.5 $\mu$m to about 2 $\mu$m, where the absorption will not critically affect most applications \cite{Osinsky2002}.  This technique has been demonstrated primarily for silicon oxynitride (SiON) films \cite{Osinsky2002,johnson2003use,Moss2014}, showing exceptionally low losses while retaining a modest index contrast.  Recently, deuterated SiN (SiN:D) has been reported with promising material performance (loss of 0.47 dB per cm at $\lambda$ = 1550 nm) for passive device applications, although the films were silicon deficient and had a substantially reduced refractive index of \textit{n} = 1.87. \cite{Hiraki2017}.  In this work, we demonstrate the viability of SiN:D waveguides for nonlinear integrated photonics.  First, we present an optimized deposition process which results in high-index, low-loss waveguides with microresonators throughout $\lambda$ = 1500 -- 1600 nm.  Next, we design and test fully-etched microresonators and demonstrate broadband modulation-instability microcombs.  Finally, we show efficient and broadband supercontinuum generation with pulsed pumping of straight and folded waveguide paths.

\begin{figure}[t]
	\centering
	\fbox{\includegraphics[width=\linewidth]{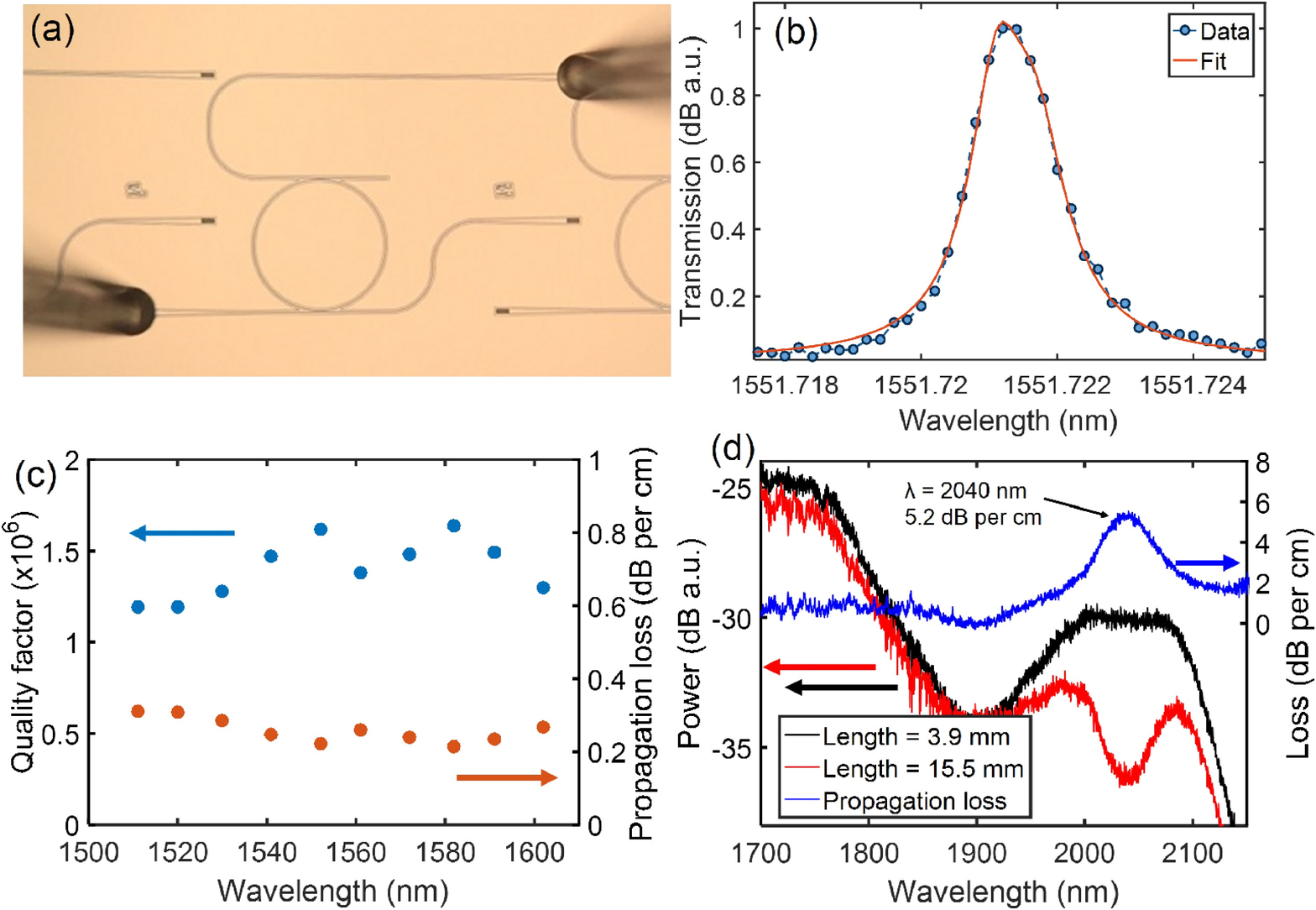}}
	\caption{(a) Optical micrograph of a shallow-etched microresonator under test with input/output optical fibers; (b) Drop-port transmission of a resonant mode at 1552 nm (the slight asymmetry is due to a weak scattering doublet); (c) Spectrum of peak total quality factor vs. wavelength; (d) Measurement of propagation loss at the N-D overtone absorption peak, showing two different device lengths (black,red) and the extracted propagation loss (blue).}
	\label{fig:testfig1}
\end{figure}

First, we describe the process for optimization of the thin film properties.  The fabrication work was performed at the Boulder Microfabrication Facility at NIST.  Prior to the installation of the deuterated precursor SiD\textsubscript{4}, a series of tests were performed with conventional SiH\textsubscript{4}, to improve the baseline SiN film quality.  Losses and refractive index were measured at 633, 1065, 1548 and 1640 nm utilizing prism coupling techniques\cite{Ulrich1973}.  Next, a gas line normally used for conventional silane gas (SiH\textsubscript{4}) was evacuated, then the tank was swapped for a cylinder containing deuterated silane (SiD\textsubscript{4}) with 98\% isotopic purity, with a total gas volume of 75 L.  This was plumbed to an inductively-coupled plasma chemical vapor deposition system (ICP-CVD).  The best deposition parameters from the trial run were taken as a starting point for the SiN:D optimization.  The gas flows needed to be adjusted by small amounts in order to compensate for the difference in molecular mass.  The best results were obtained with the following parameters: 270$^\circ$C platen temperature, 10 mTorr chamber pressure (striking at 3 mTorr and then increasing to 10 mTorr rapidly), 1000 W ICP power, 0 W radio-frequency (RF) bias power (with 30 W temporarily used to assist in striking the plasma arc), 36 sccm SiD\textsubscript{4} flow (note that this number is specified for a mass flow controller (MFC) calibrated to conventional SiH\textsubscript{4}) and 31 sccm N\textsubscript{2} flow.  The resulting films exhibit a nominal refractive index of $1.98 \pm 0.01 $ at 1548 nm.  A low tensile stress value of 191 MPa was measured for a 568 nm thick SiN:D film.  Several different thicknesses were deposited for future tests, including a 1.05 $\mu$m thick film that exhibited no cracks.  The simplicity and robustness of this process makes it a promising approach to large-scale integration of SiN photonics.

Having achieved an acceptable starting point for film quality, we moved to passive microresonator fabrication.  A 920 nm thick SiN:D film was deposited on a 76 mm diameter thermally oxidized ($\sim$ 3 $\mu$m) silicon wafer.  Prior to the deposition, the wafer was plasma-cleaned inside the ICP-CVD chamber to improve the adhesion and bottom surface quality.  Next, we patterned add-drop microresonator test devices with a radius of 180 $\mu$m with electron-beam lithography.  The patterns were partially etched by 430 nm using CF\textsubscript{4} and O\textsubscript{2} chemistry in an ICP reactive ion etching (ICP-RIE) tool.  Finally, the resist was stripped and the wafer was coated with an SiO\textsubscript{2} cladding.  Devices were tested with fiber-to-grating couplers (see Fig. \ref{fig:testfig1}(a)) to measure the (weakly loaded) total quality factors of transverse-electric (TE) polarized resonant modes across a broad wavelength range from $\lambda$ = 1510 -- 1600 nm. A peak quality factor in the C-band of $1.6\times 10^6 $ was observed at 1552 nm [Fig. \ref{fig:testfig1}(b)], corresponding to a 2$\times$ improvement in propagation loss compared to previous SiN:D waveguides \cite{Hiraki2017}.  The quality factors and corresponding linear propagation losses for a particular resonator (ring width = 2.3 $\mu$m, gap = 2 $\mu$m) are plotted in Fig. \ref{fig:testfig1}(c).  For calculating the propagation losses via the quality factor, a group index of $n_g = 2.06$ was assumed based on mode simulations.  It is notable that the loss increases somewhat at shorter wavelengths, due to the N-H absorption.  The residual $\sim$2\% hydrogen from the limited precursor purity is the most likely cause. This is easily avoided by using higher-quality precursor gas in the future.  Despite this, the worst-case propagation loss of 0.31 dB per cm represents an 8$\times$ reduction over a span from 1510 -- 1600 nm, compared to state-of-the-art work in PECVD SiN:H films \cite{Douglas2016}.  For applications such as comb generation where threshold power scales inversely with the square of the quality factor, such an improvement represents access to new functionality in low-temperature, plasma-deposited SiN films.  In the near future, improvements to the quality factor could be realized by chemical-mechanical polishing (CMP) of the starting wafer surface as well as the top surface of the SiN film \cite{Ji2016}.  Optimization of the etch recipe will also allow for higher quality factors.

We also investigated the properties of the vibrational energy shift caused by the N-H to N-D bond substitution.  Based on prior work on deuterated thin films, it is expected to reappear at a wavelength near $\sim$2 $\mu$m \cite{Osinsky2002}.  To test this, we injected a supercontinuum light source \cite{Carlson2017aa} into waveguides of different lengths, then used the cut-back method to deduce the propagation loss from the absorption peak near 2 $\mu$m.  The results are shown in Fig. \ref{fig:testfig1}(d).  A peak propagation loss of 5.2 dB per cm is observed in a narrow band centered at a wavelength of 2.04 $\mu$m. We thus conclude that this process is highly effective in relocating the most severe losses from the 1550 to 2000 nm region, allowing mature telecom laser technology to be used.

\begin{figure}[t]
	\centering
	\fbox{\includegraphics[width=\linewidth]{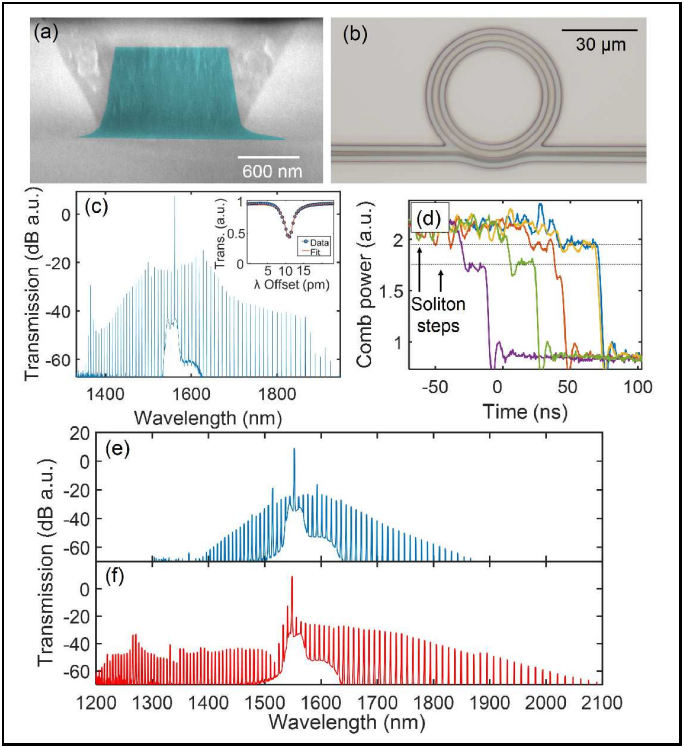}}
	\caption{(a) SEM cross section of a representative SiN:D waveguide; (b) Top-view optical micrograph of a 1 THz FSR microring resonator; (c) Comb spectrum from a 1.5 $\mu$m ring width resonator.  Inset: resonant mode used to generate this comb with total Q of $5.6\times 10^5$; (d) Oscilloscope time trace of comb power (after pump filtering) during red detuning of the pump laser frequency, showing evidence for dissipative Kerr-soliton transient formation with two discrete and repeatable step levels (dashed lines); (e-f) Comb spectra corresponding to (e) 1.6 $\mu$m and (f) 2.1 $\mu$m ring widths.}
	\label{fig:testfig2}
\end{figure}

Next, we conducted an additional fabrication run, this time focused on nonlinear photonic devices including microcomb cavities and waveguides for supercontinuum generation.  A 76 mm wafer coated with 860 nm of SiN:D was processed similarly to the previous run, except that the waveguides were fully-etched this time, and SF\textsubscript{6}+C\textsubscript{4}F\textsubscript{8} etching chemistry was utilized to achieve higher selectivity.  A scanning electron micrograph (SEM) cross section of a typical waveguide from the wafer is shown in Fig. \ref{fig:testfig2}(a). The 5 mm x 10 mm chips were released from the wafer with deep-etching to enable the simultaneous fabrication of high-quality facets on every chip.  Inverse tapers were included for every waveguide path.  A gap of 1 $\mu$m was used between the etched SiO\textsubscript{2} facet and the SiN:D taper tip.  A typical end-to-end transmission of 62\% was observed with free-space input and output coupling for a 4.6 mm long straight waveguide, corresponding to a maximum per-facet insertion loss of $1.0 \pm 0.1$ dB at $\lambda$ = 1560 nm (TE-polarized case).  Depending on the actual propagation loss of the waveguides (higher than that of a microring, due to the field stitching errors it encounters along the path), the actual per-facet insertion loss is likely lower.

We now consider the microcomb results from this fabrication run. We fabricated a separate chip with 1 THz free-spectral-range (FSR) microresonators (ring radius of 23 $\mu$m), shown in Fig. \ref{fig:testfig2}(b).  TE-polarized light was coupled on- and off-chip via lensed fibers, having a higher insertion loss of $\sim$2 dB per facet.  Several microresonators with different ring waveguide widths (\textit{W}) and coupling gaps were tested to ascertain ideal coupling conditions.  All devices employed weakly-tapered pulley couplers \cite{Spencer2014} with a coupling length of $\sim$15 $\mu$m, and 700 nm-wide bus waveguides.  A parametric oscillation threshold as low as 40 mW (estimated coupled pump power on-chip) was observed.  During testing, modulation-instability (MI) comb spectra were observed for a variety of devices, two of which are highlighted here.  The first is a resonator with a ring width of 1.5 $\mu$m and a coupling gap of 0.75 $\mu$m.  The selected mode at 1558 nm is depicted in the inset of Fig. \ref{fig:testfig2}(c).  The MI comb spectra at a continuous-wave (CW) on-chip pump power of approximately 440 mW is shown in Fig. \ref{fig:testfig2}(c), extending from 1350 to 1900 nm.  Other spectra corresponding to different ring widths are also displayed in Fig. \ref{fig:testfig2}(e-f).  In the case of \ref{fig:testfig2}(f), the spectrum is strongly shaped by the pulley coupler efficiency.  During slow laser frequency tuning (via a piezo mechanism) across the resonance, various soliton steps were observed (Fig. \ref{fig:testfig2}(d)) with a duration of approximately 20 ns.  Stable soliton generation in these devices has not yet been confirmed, due to the relatively strong thermal shift that occurs from the larger absorption coefficient (compared to annealed LPCVD SiN films), making it difficult to stabilize the spectrum when it collapses into a soliton state \cite{Li2017}.  However, the SiN:D material absorption is most likely not minimized yet, given the brief time window available to conduct these film depositions, and the large parameter space over which to search. As was the case for LPCVD SiN, we expect the properties of SiN:D to improve considerably with additional time and effort.
\begin{figure}[t]
	\centering
	\fbox{\includegraphics[width=\linewidth]{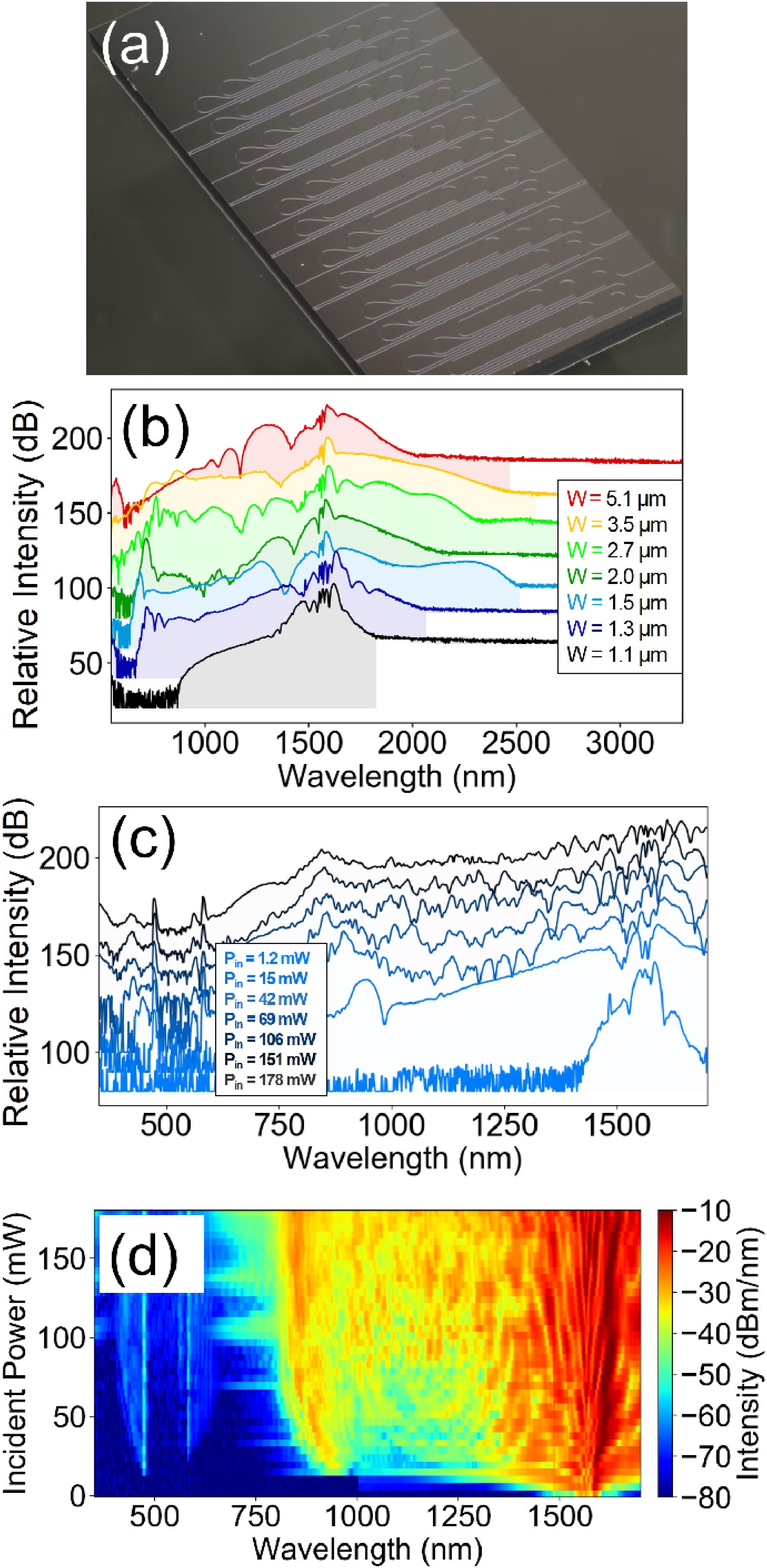}}
	\caption{(a) Focus-stacked image of a fabricated SiN:D chip for supercontinuum generation; (b) Experimentally measured output spectra for various waveguide widths with a fixed length of 3.9 mm (each line displaced by 20 dB); (c) Power scan showing the evolution of the measured output spectrum for different pump powers incident on a 3.1 $\mu$m wide waveguide (10 dB displacement per trace); (d) 2D plot of the full data from the previous power scan. Traces in (b) and (c) have an arbitrary overall offset.}
	\label{fig:testfig3}
\end{figure}

Additionally, arrays of straight and folded waveguide paths were fabricated for testing the supercontinuum generation performance (Fig. \ref{fig:testfig3}(a)).  A fiber-based mode-locked laser was used as the pump, with a center wavelength of 1560 nm, a repetition rate of 100 MHz, a pulse width of $\sim$80 fs, and an average power of up to 178 mW.  TE-polarized light was coupled on-chip via an aspheric lens (NA = 0.56), and off-chip via a multimode output fiber.  The output spectrum was resolved in either a grating-based optical spectrum analyzer (OSA) for shorter wavelengths, or a Fourier-transform spectrometer for longer wavelengths.  First, we studied the dependence of the supercontinuum spectra on the waveguide width for a constant propagation length of $\sim$4 mm.  Widths were scanned from 1.1 to 5.1 $\mu$m, and a subset of the results is shown in Fig. \ref{fig:testfig3}(b).  At \textit{W} = 1.1 $\mu$m, the waveguide experiences normal dispersion around the pump, indicated by the small degree of broadening via self phase modulation (SPM) in the spectrum.  However, anomalous-dispersion features appear at only slightly wider widths, showing a strong dispersive wave in the red or short-wave IR spectrum.  More than one octave of broadening is observed in most of the waveguide widths, despite the short propagation length.

To realize low-pulse-energy supercontinuum and dispersive wave generation, longer waveguide paths are required.  We tested paperclip-folded waveguide paths (Fig. \ref{fig:testfig3}(a)) of different lengths and waveguide widths to study the performance.  A bending radius of 160 $\mu$m was used, sufficient for relatively low bending losses at telecommunications wavelengths.  In Fig. \ref{fig:testfig3}(c-d), the evolution of the experimentally measured output spectrum with respect to increasing pump power is displayed in the case of \textit{W} = 2.6 $\mu$m and a total propagation length of 17.8 mm.  Strong dispersive wave generation at $\lambda$ = 950 nm is observed at average powers of 15 mW (corresponding to a pulse energy of 150 pJ).   

In conclusion, we fabricated high-quality deuterated silicon nitride (SiN:D) films that greatly reduce the effects of N-H bonds in the S \& C telecommunications bands, by utilizing plasma-deposition and isotopic precursor substitution.  These waveguides exhibit propagation losses below 0.31 dB per cm throughout $\lambda$ = 1510 -- 1600 nm.  Next, we fabricated waveguides and microresonators to test the nonlinear performance of the material.  Modulation-instability comb spectra were observed by pumping 1 THz FSR microrings near 1550 nm center wavelength, showing FSR-spaced comb teeth spanning from $\lambda$ = 1200 -- 2100 nm.  Additionally, long folded-path waveguides were pumped with a femtosecond laser at 1560 nm, showing supra-octave spanning supercontinuum generation, as well as low-threshold dispersive wave generation at average pump powers of $\sim$15 mW.  These promising results, as well as the simplicity of the fabrication technique, build a strong case for the continued development of the SiN:D platform.

In addition to the applications investigated in this work, we believe this platform may significantly benefit the following endeavors: (1) multi-plane (3D) photonic integration by providing back-end compatible, low-stress, low-loss and high-index-contrast waveguides \cite{Chiles2017,Sacher2017}; (2) telecommunications, by providing a path toward reliable and cost-effective SiN-based cascaded-four-wave-mixing cavities for wavelength-multiplexed transceivers \cite{Marin-Palomo2017}; (3) quantum photonics, through low-temperature processing of SiN films for entangled pair generation on pre-processed substrates \cite{Gentry2015}; and (4) straightforward heterogeneous integration of SiN waveguides with $\chi^{(2)}$ substrates such as thin-film lithium niobate on silicon \cite{Rabiei2013a} and AlGaAs-on-insulator \cite{Pu2015}.

Funding for this work is provided by NIST.

The authors thank the Boulder Microfabrication Facility staff for performing the silane cylinder swap and granting dedicated tool time to conduct the experiment. We also thank K. Srinivasan and D. Westly at NIST Gaithersburg for the SiN etching recipe, and Nathan Newbury and Ian Coddington for the use of their mode-locked laser. This work is a contribution of NIST, an agency of the U.S. government, not subject to copyright.


\bibliography{sample}

\begin{thebibliography}{10}
\newcommand{\enquote}[1]{``#1''}

\bibitem{Kippenberg2011}
T.~J. Kippenberg, R.~Holzwarth, and S.~A. Diddams,
  \enquote{{Microresonator-Based Optical Frequency Combs},}
  {\protect\JournalTitle{Science}} \textbf{332}, 555 LP -- 559 (2011).

\bibitem{Moss2013}
D.~J. Moss \emph{et~al.}, \enquote{{New CMOS-compatible platforms based on
  silicon nitride and Hydex for nonlinear optics},}
  {\protect\JournalTitle{Nature Photonics}} \textbf{7}, 597--607 (2013).

\bibitem{Brasch2016}
V.~Brasch \emph{et~al.}, \enquote{{Photonic chip-based optical frequency comb
  using soliton Cherenkov radiation.}} {\protect\JournalTitle{Science (New
  York, N.Y.)}} \textbf{351}, 357--60 (2016).

\bibitem{Liu2014}
Y.~Liu \emph{et~al.}, \enquote{{Investigation of mode coupling in
  normal-dispersion silicon nitride microresonators for Kerr frequency comb
  generation},} {\protect\JournalTitle{Optica}} \textbf{1}, 137 (2014).

\bibitem{Ji2016}
X.~Ji \emph{et~al.}, \enquote{{Breaking the Loss Limitation of On-chip
  High-confinement Resonators},} {\protect\JournalTitle{arXiv:1609.08699}}
  (2016).

\bibitem{Carlson2017}
D.~R. Carlson \emph{et~al.}, \enquote{{Self-referenced frequency combs using
  high-efficiency silicon-nitride waveguides},} {\protect\JournalTitle{Optics
  Letters}} \textbf{42}, 2314 (2017).

\bibitem{Luke2015}
K.~Luke \emph{et~al.}, \enquote{{Broadband mid-infrared frequency comb
  generation in a Si3N4 microresonator},} {\protect\JournalTitle{Optics
  Letters}} \textbf{40}, 4823 (2015).

\bibitem{Luke2013}
K.~Luke \emph{et~al.}, \enquote{{Overcoming Si3N4 film stress limitations for
  high quality factor ring resonators},} {\protect\JournalTitle{Optics
  Express}} \textbf{21}, 22829 (2013).

\bibitem{Pfeiffer:16}
M.~H.~P. Pfeiffer \emph{et~al.}, \enquote{{Photonic Damascene process for
  integrated high-Q microresonator based nonlinear photonics},}
  {\protect\JournalTitle{Optica}} \textbf{3}, 20--25 (2016).

\bibitem{Douglas2016}
E.~A. Douglas \emph{et~al.}, \enquote{{Effect of precursors on propagation loss
  for plasma-enhanced chemical vapor deposition of SiN{\_}x:H waveguides},}
  {\protect\JournalTitle{Optical Materials Express}} \textbf{6}, 2892 (2016).

\bibitem{Rabiei2013a}
P.~Rabiei \emph{et~al.}, \enquote{{Heterogeneous lithium niobate photonics on
  silicon substrates.}} {\protect\JournalTitle{Optics Express}} \textbf{21},
  25573--81 (2013).

\bibitem{Chang2017}
L.~Chang \emph{et~al.}, \enquote{{Heterogeneous integration of lithium niobate
  and silicon nitride waveguides for wafer-scale photonic integrated circuits
  on silicon},} {\protect\JournalTitle{Optics Letters}} \textbf{42}, 803
  (2017).

\bibitem{Mao2008}
S.~C. Mao \emph{et~al.}, \enquote{{Low propagation loss SiN optical waveguide
  prepared by optimal low-hydrogen module},} {\protect\JournalTitle{Optics
  Express}} \textbf{16}, 20809 (2008).

\bibitem{Shao2016}
Z.~Shao \emph{et~al.}, \enquote{{Ultra-low temperature silicon nitride photonic
  integration platform},} {\protect\JournalTitle{Optics Express}} \textbf{24},
  1865 (2016).

\bibitem{Ooi2017}
K.~J.~A. Ooi \emph{et~al.}, \enquote{{Pushing the limits of CMOS optical
  parametric amplifiers with USRN:Si7N3 above the two-photon absorption edge},}
  {\protect\JournalTitle{Nature Communications}} \textbf{8}, 13878 (2017).

\bibitem{Beyeler2001}
R.~Beyeler \emph{et~al.}, \enquote{{Material for SiON-optical waveguides and
  method for fabricating such waveguides},} {\protect\JournalTitle{WIPO
  WO2001064594 A1}}  (2001).

\bibitem{Osinsky2002}
A.~V. Osinsky \emph{et~al.}, \enquote{{Optical loss mechanisms in GeSiON planar
  waveguides},} {\protect\JournalTitle{Applied Physics Letters}} \textbf{81},
  2002--2004 (2002).

\bibitem{johnson2003use}
F.~G. Johnson \emph{et~al.}, \enquote{{Use of deuterated gases for the vapor
  deposition of thin films for low-loss optical devices and waveguides},}
  {\protect\JournalTitle{USPTO 6,614,977}}  (2003).

\bibitem{Moss2014}
D.~J. Moss \emph{et~al.}, \enquote{{Hydex Glass: a New CMOS Compatible Platform
  for All-Optical Photonic Chips},} {\protect\JournalTitle{arXiv:1404.5610}}
  (2014).

\bibitem{Hiraki2017}
T.~Hiraki \emph{et~al.}, \enquote{{Deuterated SiN/SiON Waveguides on Si
  Platform and Their Application to C-Band WDM Filters},}
  {\protect\JournalTitle{IEEE Photonics Journal}} \textbf{9}, 1--7 (2017).

\bibitem{Ulrich1973}
R.~Ulrich and R.~Torge, \enquote{{Measurement of Thin Film Parameters with a
  Prism Coupler},} {\protect\JournalTitle{Applied Optics}} \textbf{12}, 2901
  (1973).

\bibitem{Carlson2017aa}
D.~R. Carlson \emph{et~al.}, \enquote{{An ultrafast electro-optic light source
  with sub-cycle precision},} {\protect\JournalTitle{arXiv:1711.08429}}
  (2017).

\bibitem{Spencer2014}
D.~T. Spencer \emph{et~al.}, \enquote{{Integrated waveguide coupled Si3N4
  resonators in the ultrahigh-Q regime},} {\protect\JournalTitle{Optica}}
  \textbf{1}, 153 (2014).

\bibitem{Li2017}
Q.~Li \emph{et~al.}, \enquote{{Stably accessing octave-spanning microresonator
  frequency combs in the soliton regime},} {\protect\JournalTitle{Optica}}
  \textbf{4}, 193 (2017).

\bibitem{Chiles2017}
J.~Chiles \emph{et~al.}, \enquote{{Multi-planar amorphous silicon photonics
  with compact interplanar couplers, cross talk mitigation, and low crossing
  loss},} {\protect\JournalTitle{APL Photonics}} \textbf{2}, 116101 (2017).

\bibitem{Sacher2017}
W.~D. Sacher \emph{et~al.}, \enquote{{Tri-layer silicon nitride-on-silicon
  photonic platform for ultra-low-loss crossings and interlayer transitions},}
  {\protect\JournalTitle{Optics Express}} \textbf{25}, 30862 (2017).

\bibitem{Marin-Palomo2017}
P.~Marin-Palomo \emph{et~al.}, \enquote{{Microresonator-based solitons for
  massively parallel coherent optical communications},}
  {\protect\JournalTitle{Nature}} \textbf{546}, 274--279 (2017).

\bibitem{Gentry2015}
C.~M. Gentry \emph{et~al.}, \enquote{{Quantum-correlated photon pairs generated
  in a commercial 45 nm complementary metal-oxide semiconductor microelectronic
  chip},} {\protect\JournalTitle{Optica}} \textbf{2}, 1065 (2015).

\bibitem{Pu2015}
M.~Pu \emph{et~al.}, \enquote{{AlGaAs-On-Insulator Nanowire with 750 nm FWM
  Bandwidth, -9 dB CW Conversion Efficiency, and Ultrafast Operation Enabling
  Record Tbaud Wavelength Conversion},} {\protect\JournalTitle{Optical Fiber
  Communication Conference}} p. Th5A.3 (2015).

\end{thebibliography}

\bibliographyfullrefs{sample}



\end{document}